\documentclass[prc,aps,amsmath,amssymb,nofootinbib,superscriptaddress,12pt]{revtex4-2}
\usepackage{epsfig}
\usepackage[bookmarksnumbered,bookmarksopen,colorlinks,citecolor=blue,linkcolor=blue]{hyperref}
\usepackage{amsmath}
\usepackage{amssymb}
\usepackage{tipa}

\begin{document}

    \title{ Universal Wong formula for capture cross sections from light to super-heavy systems }

\author{Ning Wang}
\email{wangning@gxnu.edu.cn}\affiliation{ Guangxi Normal University, Guilin 541004, People's Republic of
	China }
\affiliation{ Guangxi Key Laboratory of Nuclear Physics and Technology, Guilin 541004, People's Republic of
	China }
 
\author{Jinming Chen}
\affiliation{  Guangxi Normal University, Guilin 541004, People's Republic of
	China }
	
\author{Yicheng Wang}
\affiliation{  Guangxi Normal University, Guilin 541004, People's Republic of
	China }
	
\author{Hong Yao}
\email{yaohong@gxnu.edu.cn}
\affiliation{  Guangxi Normal University, Guilin 541004, People's Republic of
		China }

    \begin{abstract}
        A universal Wong formula is proposed with refined model parameters for a systematic description of the capture cross sections for heavy-ion fusion reactions from  C+C to  Ni+U, in which the barrier parameters and the barrier distribution are determined by the entrance-channel nucleus-nucleus potential  based on the Skyrme energy density functional. With introducing a constraint to the width of the barrier distribution and a pocket-depth dependent barrier radius, the capture excitation functions for a number of fusion reactions involving different nuclear structure effects are remarkably well reproduced, particularly for the reactions between light nuclei and those forming super-heavy nuclei. The systematic decreasing behavior of the geometric radii with the depth of capture pocket due to the influence of deep inelastic scattering is clearly observed in the TDHF calculations for super-heavy systems. The predicted capture cross sections for $^{54}$Cr + $^{238}$U at above barrier energies are evidently smaller than the corresponding results of more asymmetric projectile-target combination $^{50}$Ti + $^{242}$Pu due to the shallower capture pocket in Cr+U.    
    \end{abstract}

     \maketitle

\newpage
  
   \begin{center}
  	\textbf{ I. INTRODUCTION }\\
  \end{center}
  
  The accurate calculations of the capture cross sections for heavy-ion fusion reactions are quite important not only for the synthesis of new super-heavy nuclei (SHN) \cite{Hof00,Hof04,Mori04a,Ogan10,Ogan15,Ogan22,Sob18,Mori20,Pomo18,Adam04,Wang15,Pei24}  but also for the exploration of nucleosynthesis in nuclear astrophysics \cite{Wea93,Fang17}. It is a challenge to describe the fusion excitation functions for all measured reactions by using a uniform method due to the uncertainty of nuclear potentials, since both complicated nuclear structure effects of the reaction partners and the dynamical effects in the fusion processes play key roles to the potentials and fusion cross sections. In the case of fusion reactions involving light and intermediate nuclei, approaches such as fusion coupled channel calculations \cite{Hag99,Dasso87,Das98} or empirical barrier distribution methods \cite{Zag01,SW04,liumin,Wang09,Wangbing,Jiang22} are adopted to calculate capture (fusion) cross sections. These calculations are often based on static or dynamic nuclear potentials \cite{Wong73,Bass74,Bass80,BW91,Gup92,Wen22,Wang08}. For different reactions systems, such as C+C and  Ni+U, the model parameters need to be re-adjusted  \cite{Bass74,Bass80} due to the difference of the projectile-target structure and the reaction channels involved, which results in some uncertainties in the predictions of the capture cross sections for unmeasured systems. 
  
  In addition to the static nuclear potentials, some microscopic dynamics models, such as the time-dependent Hartree-Fock (TDHF) \cite{Maru06,Guo07,Sim14,Yao24} theory and the improved quantum molecular dynamics (ImQMD) model \cite{ImQMD2014,ImQMD2014a} have also been widely adopted in the study of heavy-ion fusion reactions. Based on the Skyrme energy density functional \cite{Vau72} for describing the nuclear potential, these microscopic dynamics models can successfully reproduce the capture cross sections for a series of reactions at energies above the Coulomb barriers. Considering that these microscopic dynamical calculations are extremely time-consuming for massive systems, development of an analytical universal cross section formula with high accuracy is still useful for the systematic study of the fusion reactions.

 In this work, we attempt to propose a universal capture cross section formula based on Wong formula \cite{Wong73,Short} together with the Skyrme energy density functional.  The structure of this paper is as follows: In Sec. II, the frameworks of the universal Wong formula will be introduced. In Sec. III, some model parameters are refined in order to extend the formula for describing the capture cross sections from light to super-heavy systems. Simultaneously, the results from the proposed formula for a series of reaction systems are presented.  Finally a summary is given in Sec. IV.

 \begin{center}
	\textbf{ II. UNIVERSAL WONG FORMULA }\\
\end{center}

 According to Wong formula \cite{Wong73}, the fusion excitation
 function for penetrating a parabolic barrier can be expressed as,
 \begin{equation}
 	\sigma^{\rm Wong}(E,B)=\frac{\hbar \omega}{2E}  R_{m}^{2} \ln \left( 1+\exp\left[ \frac{2\pi
 	}{\hbar \omega}(E-B)\right] \right).
 \end{equation}
 Where $E$ denotes the center-of-mass incident energy. $B$, $R_m$ and $\hbar\omega$
 are the barrier height, radius and curvature, respectively.
As an one-dimensional barrier penetration model, the Wong formula is successful in
 describing the fusion excitation functions for light systems. For fusion reactions with heavy nuclei, it is known that the
 coupling of other degrees of freedom (such as deformation and
 vibration of nuclei) to the distance between two nuclei is
 obvious. Considering the multi-dimensional character of
 the realistic barriers due to the coupling to internal degrees of
 freedom of the binary system, the fixed barrier height in the traditional Wong formula $\sigma
 ^{\rm Wong}$ could be replaced by a distribution of 
 barrier heights $D(B)$. The universal Wong formula \cite{liumin},
 \begin{eqnarray}
 	\sigma _{\rm cap}(E)=\int_{0}^{ \infty }D(B) \; \sigma
 	^{\rm Wong} (E,B)dB,
 \end{eqnarray}
  is therefore proposed for describing the capture cross sections from light to heavy systems.

In Ref. \cite{liumin}, the code FUSION-v1 is proposed based on the universal Wong formula for a systematic describing the capture cross sections in heavy-ion fusion reactions, in which the barrier parameters $R_m$, $\hbar\omega$ and the distribution function $D(B)$ are determined by the entrance-channel nucleus-nucleus potential $V(R)$ \cite{liumin,Deni02} based on the Skyrme energy density functional  under frozen approximation for densities and the extended Thomas-Fermi (ETF2) \cite{Bart02} approach for  the kinetic energy density and the spin-orbit density. The distribution function $D(B)$ is expressed as a superposition
 of two Gaussian functions $D_1(B)$ and $D_2(B)$,  
 \begin{eqnarray}
 	D_{1}(B)=\frac{1}{ \sqrt{g \pi} w } \exp \left[ -\frac{%
 		(B-B_{1})^{2}}{g  w^{2}}\right]
 \end{eqnarray}
 and
 \begin{eqnarray}
 	D_{2}(B)=\frac{1}{2\sqrt{\pi}w }\exp \left[ -\frac{%
 		(B-B_{2})^{2}}{(2w )^{2}}\right].
 \end{eqnarray}
In the realistic calculations, the capture cross section is written as,
\begin{eqnarray}
	\sigma_{\rm cap}(E)=\min [ \int D_1   \sigma^{\rm Wong}   dB,  \int  (D_1 +D_2)/2  \sigma^{\rm Wong} dB ]
\end{eqnarray}
for a better description of the data.

The centriods of the Guassian functions are set as $B_{1}=f B_{0}+w/2$ and $B_{2}=f B_{0}+w$, 
with  $w =\frac{1}{2}(1-f)B_0$. $B_0$ is the frozen barrier height determined from the entrance channel nucleus-nucleus potential $V(R)$
mentioned above and the coefficient $f=0.926$ is taken based on the parameters set SkM* \cite{Bart82}. 
From Eqs.(3) and (4), one notes that the peaks and the widths of $D_1(B)$
 and $D_2(B)$ only depend on $B_0$ except the factor $g$ in $D_{1}(B)$. The quantity $g$ in $D_1(B)$ is a
 factor which empirically takes into account the structure effects of nuclei
 and has a value of $0<g \le 2$. The larger the value of $g$ is, the
 larger the capture cross section at sub-barrier energies is. For 
 fusion reactions with non-spherical (neutron-shell open) nuclei around the $\beta$-stability line, 
 the structure factor is set as $g=1$.  For the reactions with
 neutron-shell closed nuclei or neutron-rich nuclei, 
 \begin{eqnarray}
 	g=\left[ 1-c_0 \Delta Q +
 	 (\delta_{n}^{\rm prog}+\delta_{n}^{\rm targ}) /2 \right]^{-1},
 \end{eqnarray}
 where $\Delta Q=Q-Q_0$ denotes the difference between the $Q$-value
 of the system under consideration for complete fusion and that of
 the reference system.  $c_0=0.5$ MeV$^{-1}$ for $\Delta Q<0$ cases and
 $c_0=0.1$ MeV$^{-1}$ for $\Delta Q>0$ cases.
 $\delta_{n}^{\rm proj(targ)}=1$ for neutron shell closed projectile
 (target) nucleus and $\delta_{n}^{\rm proj(targ)}=0$ for non-closed
 cases (The shell-closure effects of $^{16}$O are neglected in the calculations). In addition, we introduce a truncation for $g$ value, i.e. $0<g \le 2$.  The reference system is chosen to be the system with reference nuclei along the $\beta$-stability
 line. More precisely, the mass numbers $A_0$ of the reference nuclei are determined by the relative atomic
 masses $M_{\rm a.m.}$ of the corresponding elements in the periodic table, $A_{0} -1 <
 M_{\rm a.m.}  \le A_{0} $ (with a few exceptions which will be discussed later).

\begin{center}
	\textbf{ III. EXTENSION OF THE FORMULA FOR LIGHT AND SUPER-HEAVY SYSTEMS }\\
\end{center}
 
  \begin{figure}
	\setlength{\abovecaptionskip}{ 0.2cm}
	\includegraphics[angle=0,width=1\textwidth]{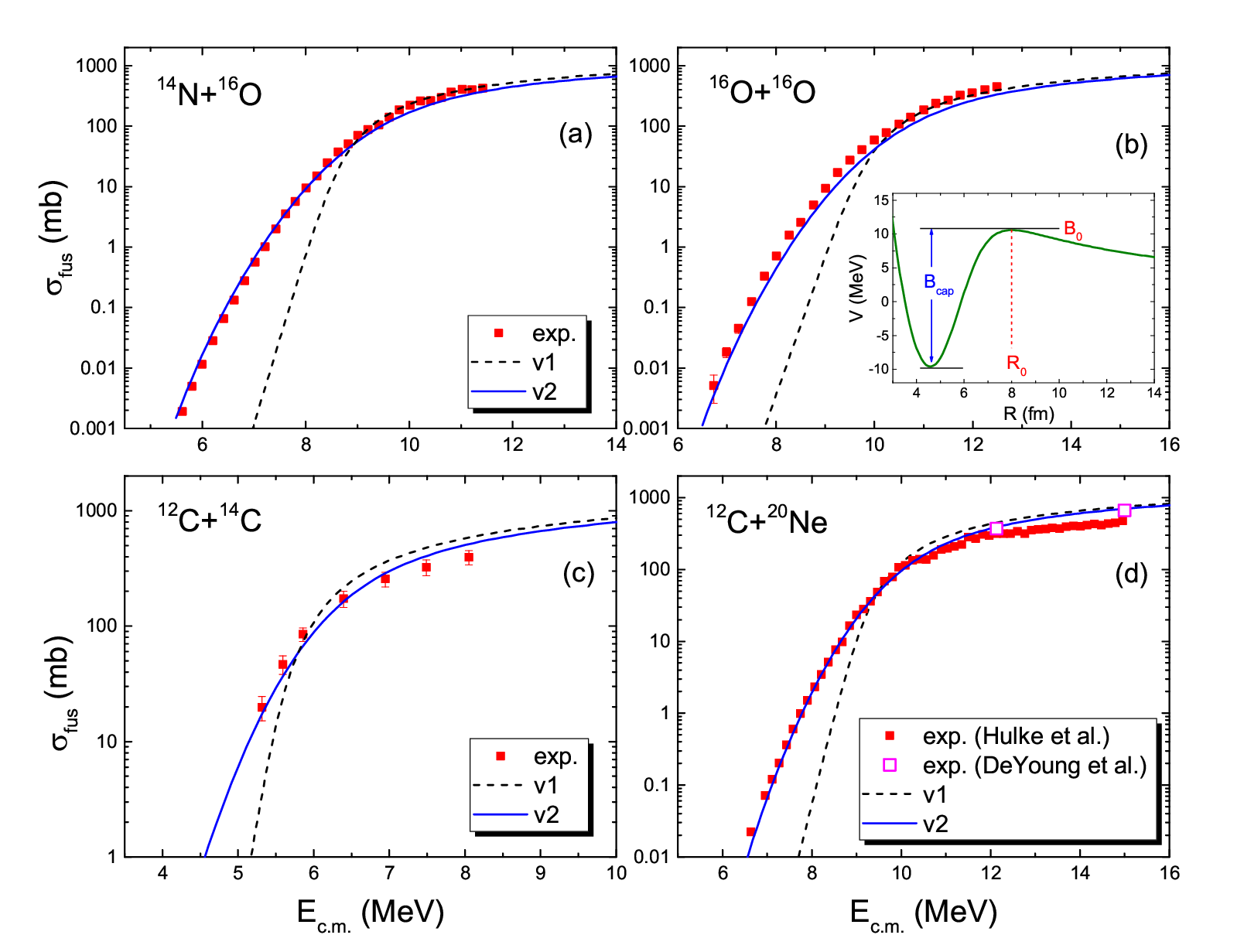}
	\caption{ Fusion excitation functions for reactions $^{14}$N + $^{16}$O \cite{Stok76}, $^{16}$O + $^{16}$O \cite{Thom85}, $^{12}$C + $^{14}$C \cite{Das93} and $^{12}$C + $^{20}$Ne \cite{Hul80,DeY82}.  Sub-figure: Entrance channel nucleus-nucleus potential for $^{16}$O + $^{16}$O. The red dashed line in the sub-figure denotes the position of the barrier radius $R_0$ and the blue line denotes the depth of the capture pocket $B_{\rm cap}$. 
	}
\end{figure}

  From the barrier distribution functions given in Eqs.(3) and (4), one notes that the width coefficient $w \propto B_0$  for all fusion reactions. In addition, we also note that the calculated fusion cross sections with FUSION-v1 for light systems such as $^{16}$O + $^{16}$O are significantly smaller than the experimental data at sub-barrier energies, although the average barrier height is close to the extracted value \cite{Chen23}. From a systematic study of the barrier parameters for 443 fusion reactions, Chen et al. find that the extracted coefficient of the distribution width  $W \approx (0.014+0.135\lambdabar_B) V_B$, in which  $V_B$ denotes the average barrier height and $\lambdabar_B$ denotes the reduced de Broglie wavelength of the colliding nuclei at an incident energy of $E=V_B$ \cite{Chen23}. For fusion reactions between light nuclei, the extracted width coefficient is obviously larger than the value of $w$ in FUSION-v1. On the other hand, it is known that the barrier distribution is smeared out with a finite width of ${\rm FWHM} \approx 0.56\hbar \omega$ in the quantum  mechanical treatment of a single parabolic potential barrier  \cite{Das98}, which is also obviously larger than the value of $w$. For a better description of the fusion cross sections for reactions between light nuclei and considering the finite width of the distribution, we add a constraint to the width coefficient $w$ in FUSION-v2, i.e., 
  \begin{eqnarray}
  	w \ge {\rm FWHM}
  \end{eqnarray}
  For fusion reactions between light well-bound nuclei,  if $w =\frac{1}{2}(1-f)B_0 < {\rm FWHM} < 1 $ MeV, we set $w = {\rm FWHM}$, $D_1=D_2$, and write the fusion cross sections as $\sigma _{\rm fus} =\int   D_2 \sigma^{\rm Wong} dB$.

In Fig. 1, we show the predicted fusion cross sections for four fusion reactions  $^{14}$N + $^{16}$O, $^{16}$O + $^{16}$O, $^{12}$C + $^{14}$C and $^{12}$C + $^{20}$Ne. The solid and the dashed curves denote the results with and without the constraint of Eq.(7) being taken into account in the calculations, respectively. 
If neglecting the constraint for $w$, one has a value of $w=0.29$ MeV  for $^{14}$N + $^{16}$O, which is significantly smaller than the corresponding finite width of ${\rm FWHM} \approx 0.56\hbar \omega=0.61$ MeV. With the constraint being considered for $^{14}$N + $^{16}$O, one sees that the experimental data can be  remarkably well reproduced. The better reproduction of the experimental data for these light systems indicate the constraint to the value of $w$ is necessary and reasonable.

 For the fusion reactions leading to the synthesis of super-heavy nuclei, the depth of the capture pocket $B_{\rm cap}$ in the entrance channel nucleus-nucleus potential $V(R)$ is much shallower than that of light systems, and the quasi-fission (QF) becomes evident. In addition, for some super-heavy systems such as $^{64}$Ni+$^{238}$U \cite{Kozu10,Itkis22}, the extracted capture cross sections from the measured mass-total kinetic energy (TKE) distributions at energies above the Bass barrier  $E_{\rm Bass}$ \cite{Bass74} are significantly smaller than the results of FUSION-v1. To understand the physics behind, we systematically study the capture cross sections for some fusion reactions with the time dependent Hartree-Fock (TDHF) calculations at energies above $E_{\rm Bass}$. It is thought that the contact time of the composite system is about 2 zs ($\sim$ 600 fm/c) for the capture process \cite{Yao24}. If the composite system reseparates into two fragments within 2 zs after projectile-target contact, we treat it as inelastic scattering rather than QF \cite{Lee24}. We calculate the contact times for the reaction systems at $E=1.05 E_{\rm Bass}$ at a certain impact parameter. If the contact time is larger than 600 fm/c, the simulation is terminated to save CPU hours. The sub-figure in Fig. 2 shows the contact time as a function of impact parameter for reactions $^{86}$Kr + $^{208}$Pb, $^{64}$Ni + $^{208}$Pb, $^{58}$Fe+ $^{208}$Pb and $^{40}$Ca + $^{96}$Zr. One sees that for a certain reaction the contact time decreases from 600 fm/c at a critical impact parameter $b_{\rm cap}$ to zero at $b_T$. With the critical impact parameter $b_{\rm cap}$, the capture cross section $\sigma_{\rm cap}=\pi b_{\rm cap}^2$ can be obtained. Simultaneously, one can obtain the touching cross section $\sigma_T=\pi b_T^2$ which approximately represents the probability of the reaction partners overcoming the Coulomb barrier. In Fig. 2, we show the calculated ratio $\sigma_{\rm cap}/ \sigma_T$ as a function of capture pocket depth $B_{\rm cap}$. The squares denote the results of TDHF for these reactions. For $^{40}$Ca + $^{96}$Zr, the ratio is about one. For heavier systems, the ratio significantly decreases with the decreasing of the pocket depth $B_{\rm cap}$. It implies that the influence of deep inelastic scattering (DIS) on the capture process becomes stronger for heavier systems producing super-heavy nuclei, which is also observed in Ref. \cite{Yao24a} (in which the pocket depth is denoted by $B_{\rm qf}$).

  \begin{figure}
 	\setlength{\abovecaptionskip}{ 0.2cm}
 	\includegraphics[angle=0,width=0.85\textwidth]{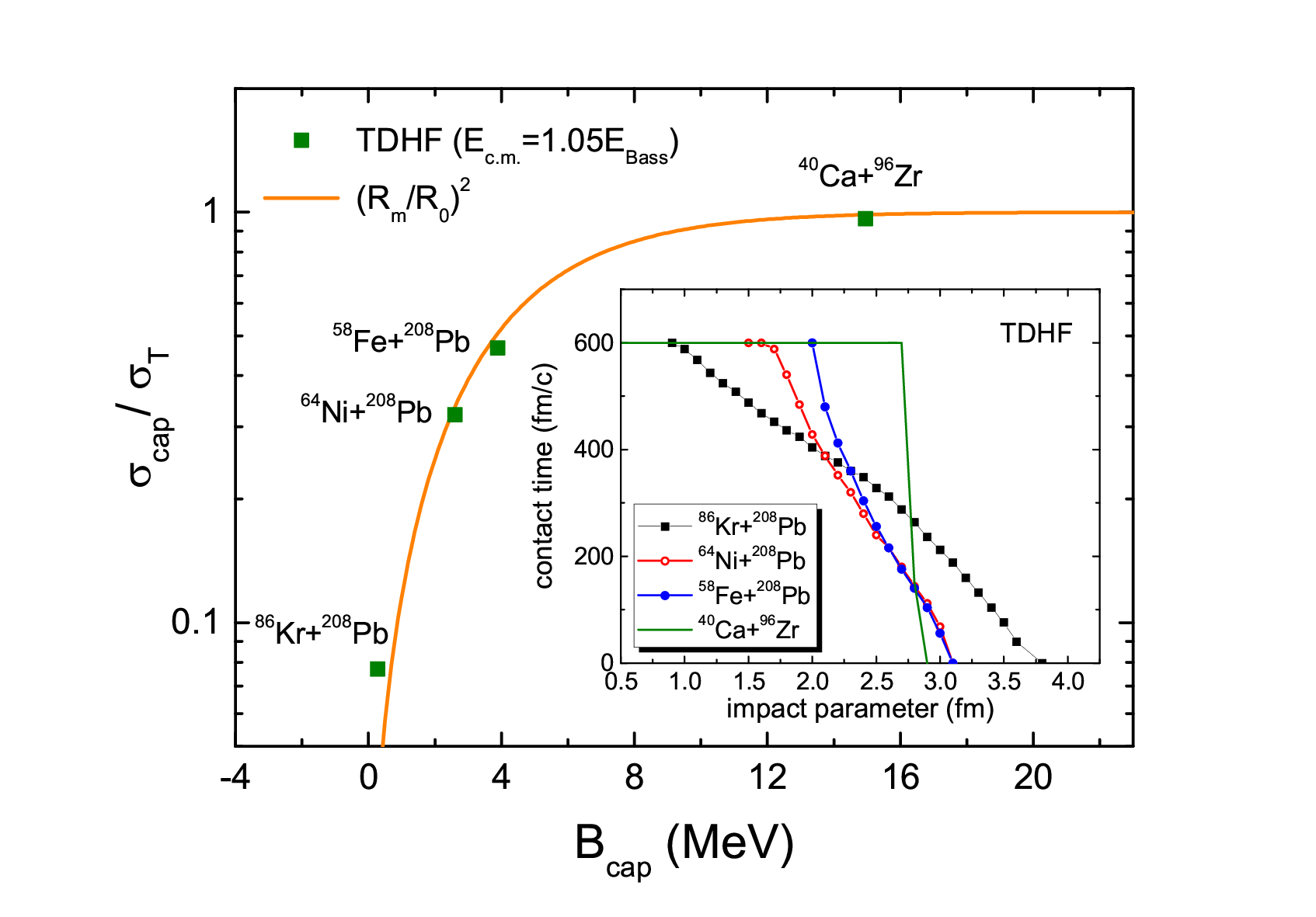}
 	\caption{  Ratio of capture cross section to touching cross section as a function of capture pocket depth. The squares and the curve denote the results from the TDHF calculations and those with Eq.(8), respectively. Sub-figure: Contact time of reaction system in the TDHF calculations as a function of impact parameter.}
 \end{figure}
 
 \begin{figure}
 	\setlength{\abovecaptionskip}{ -1 cm}
 	\includegraphics[angle=0,width=1\textwidth]{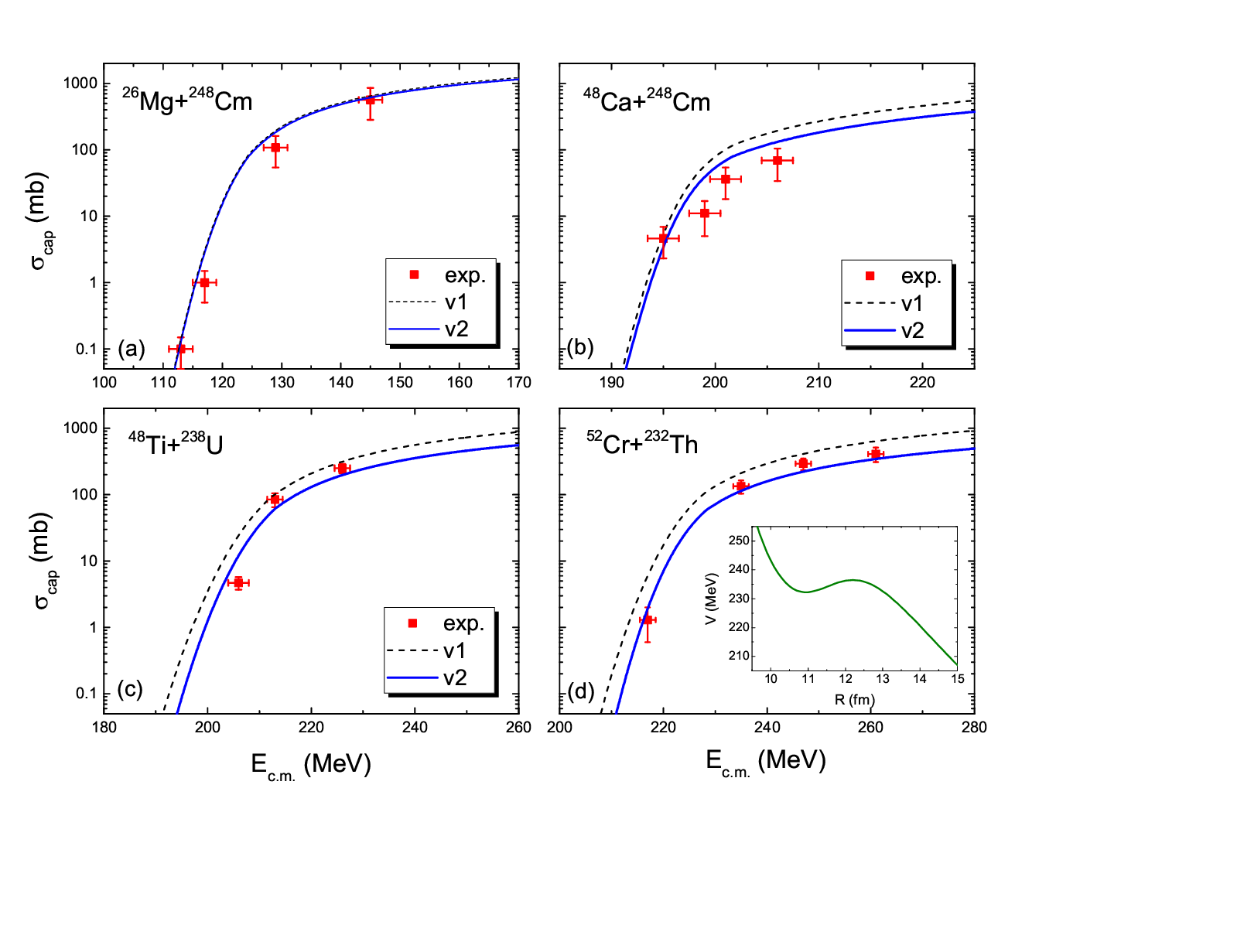}
 	\caption{ (a) The same as Fig. 1, but for reactions $^{26}$Mg + $^{248}$Cm, $^{48}$Ca + $^{248}$Cm, $^{48}$Ti + $^{238}$U and $^{52}$Cr + $^{232}$Th. The measured capture cross sections are taken from \cite{Itkis22}.  Sub-figure: Entrance channel nucleus-nucleus potential for $^{52}$Cr + $^{232}$Th. }
 \end{figure}
 
 In FUSION-v1, the influence of DIS on the capture cross sections is neglected, which probably results in the over-prediction of the measured capture cross sections for super-heavy systems such as $^{64}$Ni + $^{238}$U. For a better description of the capture cross sections for super-heavy systems, a factor $F_{\rm DIS}$ is introduced in FUSION-v2, 
     \begin{eqnarray}
   	F_{\rm DIS}= \frac{1}{2}\left [ 1+{\rm erf}(\sqrt{B_{\rm cap}/c_1}-1) \right ],
   \end{eqnarray}   
with $c_1=2.0$ MeV. The barrier radius $R_0$ in the potential $V(R)$ (see the sub-figure in Fig. 1) and the structure factor $g$ are multiplied by $F_{\rm DIS}$ in the calculations  to consider the influence of DIS. The average barrier radius is therefore written as $R_m=R_0 F_{\rm DIS}$. For light fusion systems 	$F_{\rm DIS} \simeq 1$ due to the deep capture pocket. The ratio  $\sigma_{\rm cap}/ \sigma_T = (R_m/R_0)^2$ according to the classic cross section formula. The solid curve in Fig. 2 shows the calculated ratios $\sigma_{\rm cap}/ \sigma_T$  with Eq.(8). One can see that the values of $(R_m/R_0)^2$ with Eq.(8) are in good agreement with the results from the TDHF calculations. For light fusion system, the compound nucleus would be directly formed after the capture barrier being overcome due to the deep capture pocket, and therefore $R_m \simeq R_0$, $\sigma_{\rm cap}\simeq \sigma_T$ hold.  In Fig. 3, we compare the predicted capture excitation functions with and without Eq.(8) being taken into account in the calculations. With Eq.(8) for describing the average barrier radius, the experimental data are better reproduced, especially for the systems with heavier projectile nuclei. The sub-figure shows the entrance channel nucleus-nucleus potential $V(R) $ for $^{52}$Cr + $^{232}$Th. One can see that the depth of the capture pocket is only about 4.2 MeV, which is much smaller than that of $^{16}$O + $^{16}$O in Fig. 1. The shallow capture pocket leads to the reduction of the capture cross sections at energies above the barrier.

\begin{figure}
	\setlength{\abovecaptionskip}{ 0.2 cm}
	\includegraphics[angle=0,width=1\textwidth]{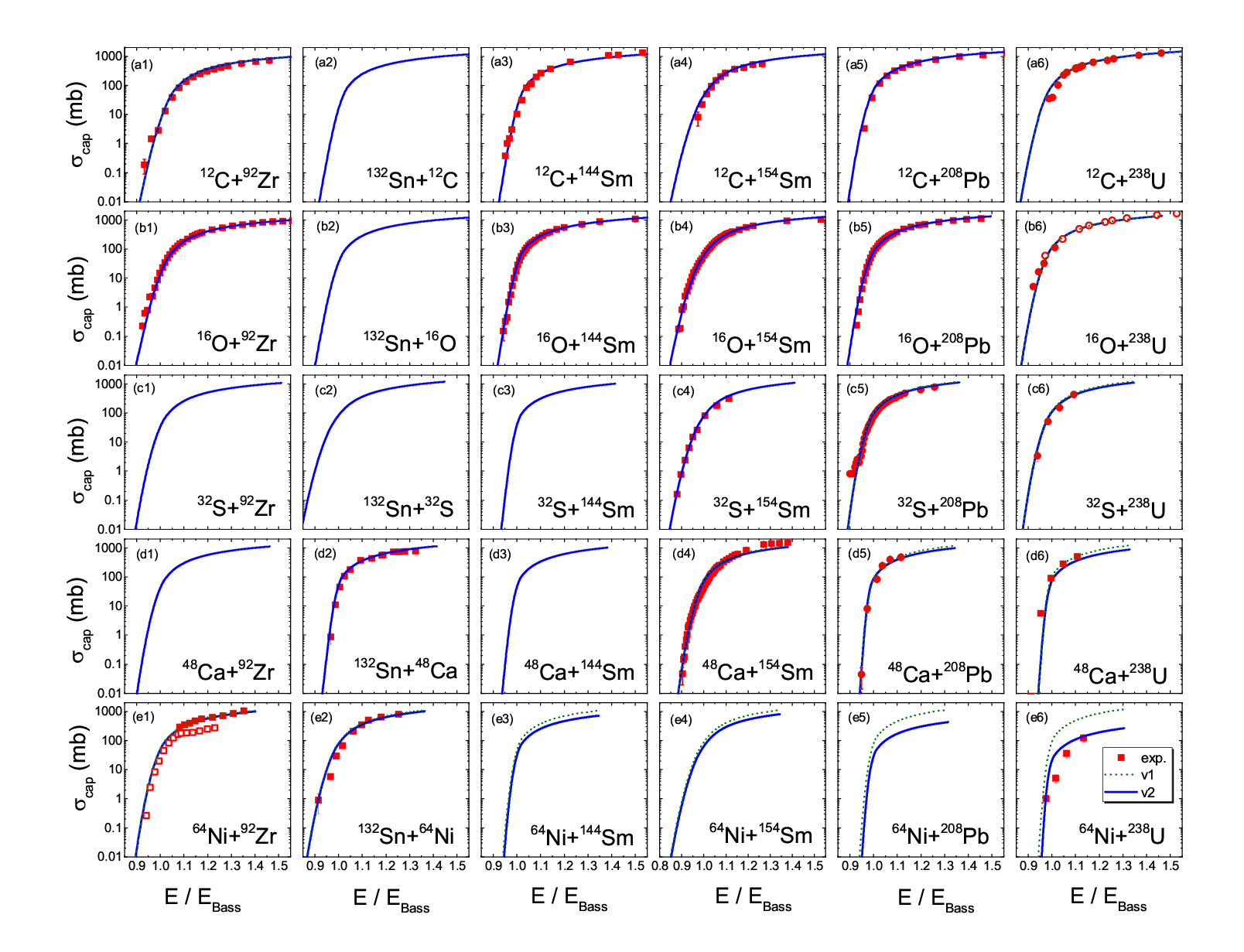}
	\caption{ Capture excitation functions for fusion reactions with $^{12}$C, $^{16}$O, $^{32}$S, $^{48}$Ca and $^{64}$Ni bombarding severally on $^{92}$Zr, $^{132}$Sn, $^{144}$Sm, $^{154}$Sm,  $^{208}$Pb and $^{238}$U. The squares and circles denote the measured capture cross sections and fusion-fission cross sections, respectively, which are taken from \cite{New01,Abr92,Gil85,Muk07,Vio62,Lei95,Zhang94,Gom94,Hind07,Frei87,Kola12,Trot05,Prok08,Stef90,Wolf89,Liang07,Itkis22}. The dashed and the solid curves denote the results with v1 and v2, respectively.  Here, the incident energy is scaled by the Bass barrier $E_{\rm Bass}$ \cite{Bass74}. }
\end{figure}

 Another modification in FUSION-v2 is that the reference systems for the reactions induced by lanthanides are refined. In FUSION-v1,  the reference system needs to be determined by measured cross sections for the reactions induced by lanthanides due to the large deformations of nuclei, which results in some uncertainties of the structure factor $g$ in the calculations for unmeasured reactions. In FUSION-v2, the mass numbers of the reference lanthanides are set as $(A_0+A_0^\prime)/2$. $A_0$ denotes the mass number determined by the relative atomic masses $M_{\rm a.m.}$ (i.e., $A_{0} -1 < M_{\rm a.m.}  \le A_{0} $) as mentioned previously. $A_0^\prime$ denotes the mass number of the lightest stable isotope of the corresponding element.

To test the model accuracy of FUSION-v2, we systematically calculate the capture excitation functions for 30 fusion reactions with $^{12}$C, $^{16}$O, $^{32}$S, $^{48}$Ca and $^{64}$Ni bombarding severally on $^{92}$Zr, $^{132}$Sn, $^{144}$Sm, $^{154}$Sm,  $^{208}$Pb and $^{238}$U. In these reactions, not only the shell effect, the deformation effect, but also the isospin effect in extremely neutron-rich nuclei are involved. The predicted capture excitation functions for these reactions are shown in Fig. 4. The squares and circles denote the measured capture cross sections and fusion-fission cross sections, respectively. We would like to emphasize that for all reactions under consideration the values of the model parameters are fixed and no additional adjustable parameter is introduced in the calculations. For medium-mass fusion systems, such as the reactions induced by $^{12}$C and $^{16}$O in Fig. 4, the results of v1 and those of v2 are very close to each other. For super-heavy systems, such as $^{64}$Ni+$^{238}$U, the results of v2 are evidently smaller than those of v1 at energies above the barrier. From Fig. 4, one can see that almost all data are well reproduced with v2, which indicates the universal Wong formula is reliable for a systematic description of the capture cross sections from light to super-heavy systems. From a systematic comparison of the capture cross sections at energies above the Bass barriers for $^{238}$U induced reactions, one could note that the capture cross sections decrease from more than 1000 mb for $^{12}$C+$^{238}$U to a few hundreds millibarn for $^{64}$Ni+$^{238}$U, although the geometric radius is much larger for the latter. This trend is also clearly observed by Kozulin et al. in experiments \cite{Koz16}. In addition, the systematic decreasing behavior of the geometric radii with effective fissility parameter is also observed from 443 datasets of measured cross sections \cite{Chen23}. It indicates that the influence of deep inelastic scattering for super-heavy systems needs to be considered in the calculations.

   \begin{figure}
 	\setlength{\abovecaptionskip}{ 0.2cm}
 	\includegraphics[angle=0,width=0.85\textwidth]{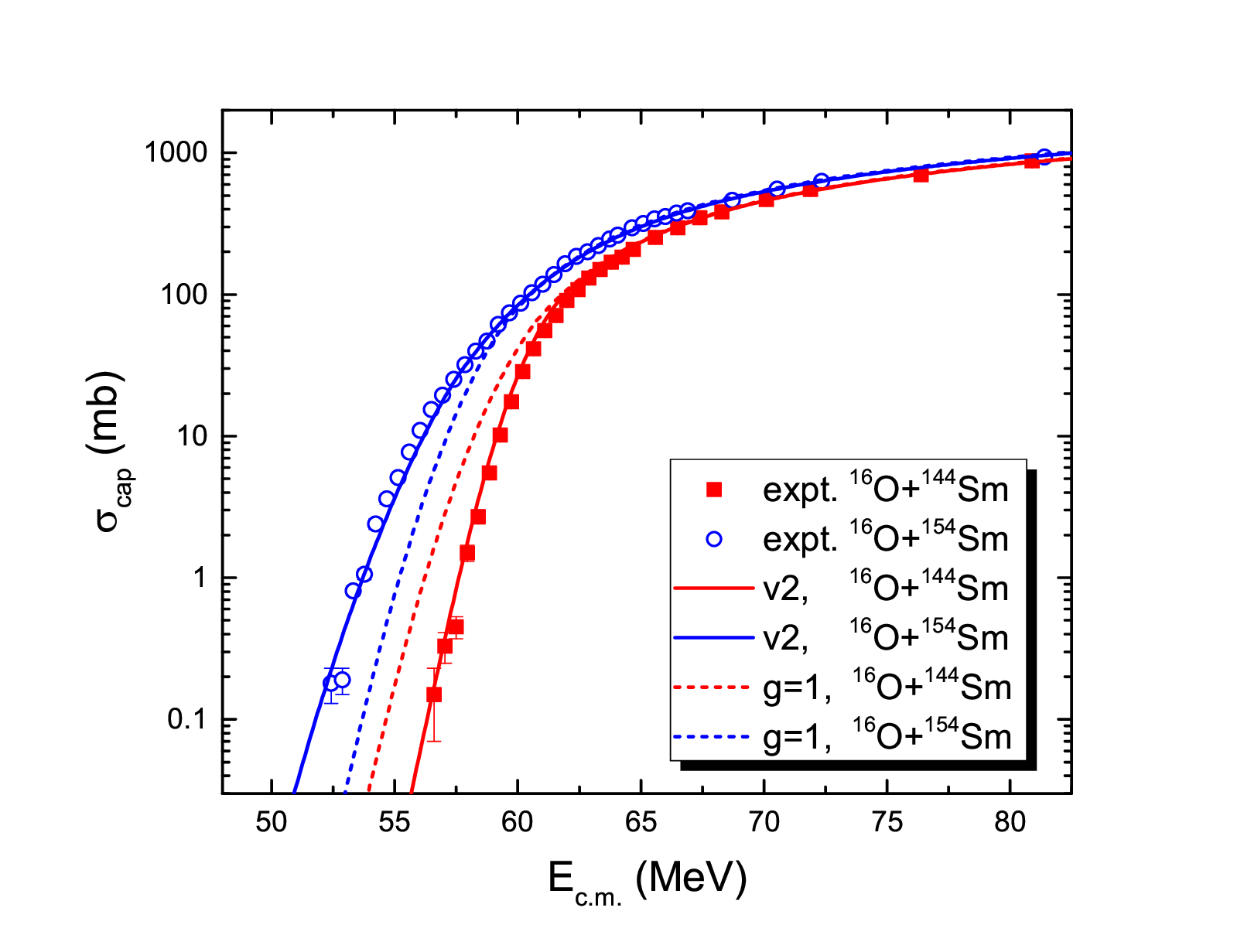}
 	\caption{ Capture excitation functions for reactions  $^{16}$O + $^{144,154}$Sm. The squares and circles denote the measured fusion cross sections \cite{Lei95} for  $^{16}$O + $^{144}$Sm and  $^{16}$O + $^{154}$Sm, respectively. The solid curves denote the predicted results with FUSION-v2. The short-dashed curves denote the results neglecting the influence of structure effects and taking $g=1$.
 	}
 \end{figure}
   
 To see the influence of nuclear structure effects on capture cross sections, we analyze the fusion reactions $^{16}$O + $^{144}$Sm and  $^{16}$O + $^{154}$Sm. $^{144}$Sm is nearly spherical in shapes due to the neutron shell closure and $^{154}$Sm is well deformed. Fig. 5 shows the predicted capture excitation functions for $^{16}$O + $^{144,154}$Sm. The dashed curves denote the results neglecting the influence of structure effects and taking $g=1$. One sees that if taking $g=1$, the measured data for $^{16}$O + $^{144}$Sm are over-predicted and the data for $^{16}$O + $^{154}$Sm are under-predicted at sub-barrier energies. Considering the structure effects of target nuclei, the obtained values of structure factor according to Eq.(6) are $g=0.29$ and $2.00$ for $^{16}$O + $^{144}$Sm and  $^{16}$O + $^{154}$Sm, respectively. Considering nuclear structure effects in the universal Wong formula, the measured cross sections for these two reactions can be remarkably well reproduced. Here, we would like to emphasize that the present version of the universal Wong formula is based on the parabolic barrier assumption, which could not be applicable for deep sub-barrier fusion since the quick increase of the barrier width due to Coulomb interaction and the repulsive of nuclear force \cite{Mis07,Nhu21} would play a role at extreme sub-barrier energies.

 \begin{figure}
 	\setlength{\abovecaptionskip}{ -1.5 cm}
 	\includegraphics[angle=0,width=0.95\textwidth]{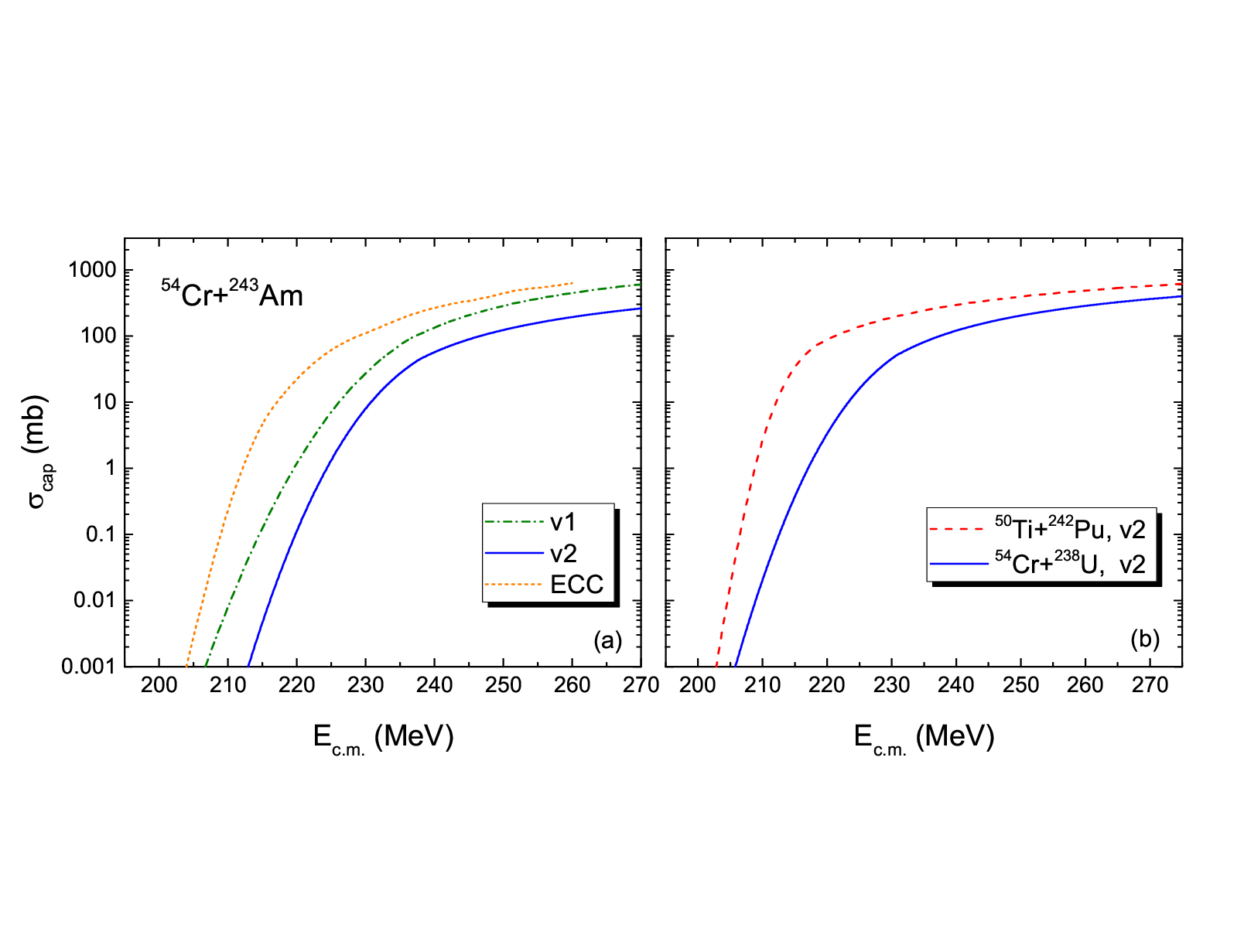}
 	\caption{ Predicted capture excitation functions for fusion reactions $^{54}$Cr + $^{243}$Am, $^{54}$Cr + $^{238}$U and $^{50}$Ti + $^{242}$Pu. The short dashed and the dot-dashed curves in (a) denote the results of empirical coupled channel (ECC) approach \cite{Wangbing,wangbing21} and those with FUSION-v1, respectively. }
 \end{figure}
 
In Fig. 6, we show the predicted capture cross sections for  $^{54}$Cr + $^{243}$Am, $^{54}$Cr + $^{238}$U and $^{50}$Ti + $^{242}$Pu. From Fig. 5(a), one can see that the predicted capture cross sections with FUSION-v2 are significantly smaller than those with FUSION-v1 for  $^{54}$Cr + $^{243}$Am due to the influence of deep inelastic scattering, and the result from the empirical coupled channel (ECC) approach \cite{Wangbing,wangbing21} is higher than that of v2 by about a factor of 7 at an incident energy of $E_{\rm c.m.}=235$ MeV. From Fig. 5(b), one notes that the predicted capture cross sections for $^{54}$Cr + $^{238}$U at above barrier energies are evidently smaller than the corresponding results of more asymmetric fusion system $^{50}$Ti + $^{242}$Pu, due to the shallower capture pocket in Cr+U ($B_{\rm cap}=3.80$ MeV for $^{54}$Cr + $^{238}$U and $B_{\rm cap}=4.58$ MeV for $^{50}$Ti + $^{242}$Pu). The smaller capture cross sections for $^{54}$Cr + $^{238}$U may result in smaller  evaporation residual cross sections considering that the same compound nucleus is formed in $^{54}$Cr + $^{238}$U and $^{50}$Ti + $^{242}$Pu. Very recently, the  evaporation residue cross sections for these two reactions have already been measured \cite{Ogan24}. The observed cross section of $^{54}$Cr + $^{238}$U is much smaller than that of $^{50}$Ti + $^{242}$Pu as expected.  We also compared the predicted capture cross sections from  KEWPIE2-EBD \cite{EBD} (which is based on an assumption that fusion barriers are normally distributed around a mean value \cite{SW04}) and those of the universal Wong formula (FUSION-v2) for the fusion reactions $^{9}$Be + $^{238}$U \cite{Mei}, $^{30}$Si + $^{238}$U, $^{40}$Ca + $^{238}$U and $^{54}$Cr + $^{243}$Am. We note that the results of EBD and FUSION-v2 are close to each other for $^{9}$Be + $^{238}$U and the experimental data can be reproduced reasonably well. For $^{30}$Si + $^{238}$U and $^{40}$Ca + $^{238}$U, the data are better reproduced by FUSION-v2. For  $^{54}$Cr + $^{243}$Am, the results of EBD are close to those of FUSION-v2 at sub-barrier energies. At energies above the capture barrier by $\sim$10\%, the results of EBD are higher than those of FUSION-v2 by a factor of two, since the influence of deep inelastic scattering on capture is neglected in EBD calculations for superheavy systems.
  
\begin{center}
	\textbf{IV. SUMMARY}
\end{center}

Based on the frozen nucleus-nucleus potential from the Skyrme energy density functional together with a barrier distribution composed of a combination of two Gaussian  functions to account for the dynamic effects in fusion processes, a universal Wong formula is proposed for a systematic description of the capture cross sections from light to super-heavy systems. With introducing a constraint to the width of the barrier distribution and a pocket-depth dependent barrier radius, the capture excitation functions for a number of fusion reactions involving different nuclear structure effects are well reproduced, particularly for the light systems such as $^{12}$C+$^{14}$C, $^{16}$O+$^{16}$O and the massive systems such as $^{52}$Cr + $^{232}$Th, $^{64}$Ni + $^{238}$U. For super-heavy systems, the systematic decreasing behavior of the geometric radii with the depth of capture pocket can be clearly observed in the TDHF calculations, which indicates the influence of deep inelastic scattering  needs to be considered for a reliable description of the capture cross sections in the synthesis of new super-heavy nuclei. With the proposed universal Wong formula for describing the capture cross sections,  the evaporation residual cross sections for fusion reactions leading to the synthesis of super-heavy nuclei could be further investigated with less uncertainties. We note that the predicted capture cross sections for $^{54}$Cr + $^{238}$U at above barrier energies are evidently smaller than the corresponding results of more asymmetric projectile-target combination $^{50}$Ti + $^{242}$Pu due to the shallower capture pocket in Cr+U, which is consistent with the trend of the measured evaporation residue cross sections.

\begin{center}
	\textbf{ACKNOWLEDGEMENTS}
\end{center}
This work was supported by National Natural Science Foundation of
China (Nos. 12265006, 12375129, U1867212), Guangxi Natural Science Foundation (2017GXNSFGA198001) and Innovation Project of Guangxi Graduate Education (YCSWYCSW2022176). The codes FUSION for calculating capture cross sections are available at http://www.imqmd.com/fusion/


\begin{thebibliography}{99}
	
	
\bibitem{Hof00} S. Hofmann and G. M\"unzenberg, 
Rev. Mod. Phys. \textbf{72}, 733  (2000). 

\bibitem{Hof04} S. Hofmann, F.P. Hessberger, D. Ackermanna et al., 
Nucl. Phys. A \textbf{734}, 93  (2004). 

\bibitem{Mori04a}  K. Morita, K. Morimoto, D. Kaji et al., 
J. Phys. Soci. Japan	\textbf{73}, 2593  (2004). 


\bibitem{Ogan10} Yu. Ts. Oganessian, F.Sh. Abdullin, P.D. Bailey et al., 
Phys. Rev. Lett. \textbf{104}, 142502 (2010). 
	
\bibitem{Ogan15} Yu. Ts. Oganessian, V.K. Utyonkov, 
Nucl. Phys. A \textbf{944}, 62  (2015). 
	
\bibitem{Ogan22} Yu. Ts. Oganessian,  V.K. Utyonkov, N.D. Kovrizhnykh et al., 
Phys. Rev. C \textbf{106}, L031301 (2022). 
	
\bibitem{Sob18} A. Sobiczewski, Yu.A. Litvinov, M. Palczewski, 
Atom. Data Nucl. Data Tables \textbf{119}, 1   (2018). 
	
\bibitem{Mori20} T. Tanaka, K. Morita, K. Morimoto et al., 
Phys. Rev. Lett. \textbf{124},  052502  (2020). 
	
\bibitem{Pomo18} K. Pomorski, B. Nerlo-Pomorska, J. Bartel et al., 
Phys. Rev. C \textbf{97}, 034319  (2018). 
	
\bibitem{Adam04} G.G. Adamian, N.V. Antonenko, and W. Scheid, 
Phys. Rev. C \textbf{69},  011601(R)  (2004). 
	
\bibitem{Wang15} Y.Z. Wang, S.J. Wang, Z.Y. Hou et al., 
Phys. Rev. C \textbf{92},  064301  (2015). 

\bibitem{Pei24} D.W. Guan, J.C. Pei, 
Phys. Lett. B \textbf{851}, 138578 (2024). 
	 
\bibitem{Wea93} T. A. Weaver, S.E. Woosley, Phys. Rep. \textbf{227}, 65 (1993). 

\bibitem{Fang17} X. Fang, W. P. Tan, M. Beard, et al., Phys. Rev. C \textbf{96}, 045804 (2017). 

\bibitem{Das98} M. Dasgupta, D.J. Hinde, N. Rowley et al., 
Annu. Rev. Nucl. Part. Sci. \textbf{48}, 401  (1998). 
	  
\bibitem{Hag99} K. Hagino, N. Rowley, and A.T. Kruppa, 
Comput. Phys. Commun. \textbf{123}, 143 	  (1999). 
	  
\bibitem{Dasso87} C.H. Dasso, S. Landowne, 
Comput. Phys. Commun. \textbf{46}, 187 (1987).  
	  
	 
\bibitem{Zag01} V.I. Zagrebaev, Y. Aritomo, M.G. Itkis et al., 
 Phys. Rev. C \textbf{65}, 014607 (2001). 

\bibitem{SW04} K. Siwek-Wilczy\ifmmode~\acute{n}\else \'{n}\fi{}ska and  J. Wilczy\ifmmode~\acute{n}\else \'{n}\fi{}ski, 
Phys. Rev. C \textbf{69}, 024611 (2004). 

\bibitem{liumin} M. Liu, N. Wang, Z.X. Li et al., 
Nucl. Phys. A \textbf{768}, 80 (2006). 

\bibitem{Wang09} N. Wang, M. Liu, Y.X. Yang, 
Sci. China Ser. G - Phys. Mech. Astron. \textbf{52}, 1554 (2009). 

\bibitem{Wangbing} B. Wang, K. Wen, W.J. Zhao et al., 
At. Data Nucl. Data Tables \textbf{114}, 281 (2017). 

\bibitem{Jiang22} C.L. Jiang and B.P. Kay, 
Phys. Rev. C \textbf{105}, 064601 (2022). 


\bibitem{Wong73}  C.Y. Wong, 
Phys. Rev. Lett. \textbf{31}, 766 (1973). 

\bibitem{Bass74} R. Bass, 
Nucl. Phys. A \textbf{231}, 45 (1974). 


\bibitem{Bass80} R. Bass, 
\emph{Lecture Notes in Physics}, \textbf{117} (Berlin: Springer) 281  (1980). 


\bibitem{BW91} R.A. Broglia, A. Winther, \emph{Heavy Ion Reactions}, Frontiers
in Physics, Vol. \textbf{84}, (Addison-Wesley) (1991).

\bibitem{Gup92} Rajeev K. Puri and Raj K. Gupta, 
Phys. Rev. C \textbf{45}, 1837 (1992). 

\bibitem{Wen22} P.W. Wen, C.J. Lin, H.M. Jia et al., 
Phys. Rev. C \textbf{105}, 034606 (2022). 

\bibitem{Wang08} N. Wang, W. Scheid, 
Phys. Rev. C \textbf{78}, 014607 (2008). 


\bibitem{Maru06} J.A. Maruhn, P.G. Reinhard, P.D. Stevenson et al., 
Phys. Rev. C \textbf{74}, 027601 (2006). 

\bibitem{Guo07} Lu Guo, J.A. Maruhn, and P.G. Reinhard, 
Phys. Rev. C \textbf{76}, 014601 (2007). 

\bibitem{Sim14} C. Simenel, 
J. Phys. G: Nucl. Part. Phys. \textbf{41}, 094007 (2014); arXiv:1403.3246v1. 

\bibitem{Yao24} H. Yao, H. Yang, N. Wang, 
Phys. Rev. C \textbf{110}, 014602 (2024); 


\bibitem{ImQMD2014} N. Wang, L. Ou, Y.X. Zhang and Z.X. Li, 
Phys. Rev. C \textbf{89}, 064601 (2014). 

\bibitem{ImQMD2014a} N. Wang,  K. Zhao and Z.X. Li, 
Phys. ReV. C \textbf{90}, 054610 (2014). 


\bibitem{Vau72} D. Vautherin, D.M. Brink, 
Phys. Rev. C \textbf{5}, 626  (1972). 

\bibitem{Short} J.M.B. Shorto, P.R.S. Gomes, J. Lubian, L.F. Canto, S. Mukherjee, L.C. Chamon, Phys. Lett. B \textbf{678}, 77 (2009). 
 
\bibitem{Deni02} V. Yu. Denisov and W. Noerenberg, 
Eur. Phys. J. \textbf{A15},  375 (2002). 

\bibitem{Bart02} J. Bartel and K. Bencheikh, 
Eur. Phys. J, \textbf{A14},  179  (2002). 

\bibitem{Bart82}  J. Bartel, Ph. Quentin, M. Brack et al., 
Nucl. Phys. A \textbf{386},  79  (1982). 


\bibitem{Chen23} Y. Chen, H. Yao, M. Liu et al., 
 Atom. Data Nucl. Data Tables \textbf{154}, 101587 (2023). 





\bibitem{Stok76} R.G. Stokstad, Z. E. Switkowski, R.A. Dayras et al., Phys. Rev. Lett. \textbf{37}, 888 (1976). 

\bibitem{Thom85} J. Thomas, Y. T. Chen, S. Hinds et al.,Phys. Rev. C \textbf{31}, 1980(R) (1985). 

\bibitem{Das93} B. Dasmahapatra, B. $\breve{C}$ujec, Nucl. Phys. A \textbf{565}, 657 (1993). 

\bibitem{Hul80} G. Hulke, C. Rolfs, H. P. Trautvetter, Z. Phys. A \textbf{297}, 161 (1980).

\bibitem{DeY82} P. A. DeYoung, J. J. Kolata, R. C. Luhn et al., Phys. Rev. C \textbf{25}, 1420 (1982).



\bibitem{Kozu10}	 E.M. Kozulin, G.N. Knyazheva, I.M. Itkis et al., 
Phys. Lett. B \textbf{686},  227 (2010). 

\bibitem{Itkis22} M.G. Itkis, G.N. Knyazheva, I.M. Itkisa et al., 
Eur. Phys. J. A \textbf{58},  178 (2022). 


\bibitem{Lee24} H. Lee, P. McGlynn, and C. Simenel, Phys. Rev. C \textbf{110}, 024606 (2024). 



 \bibitem{Yao24a}	 H. Yao, C. Li,  H.B. Zhou, and N. Wang, 
 Phys. Rev. C \textbf{109}, 034608 (2024). 
 
 
 \bibitem{New01} J.O. Newton, C.R. Morton, M. Dasgupta et al., Phys. Rev. C \textbf{64}, 064608 (2001). 
 \bibitem{Abr92} D. Abriola, A.A. Sonzogni, M. di Tada et al., Phys. Rev. C \textbf{46}, 244 (1992). 
 \bibitem{Gil85} S. Gil, R. Vandenbosch, A. J. Lazzarini et al., Phys. Rev. C \textbf{31}, 1752 (1985). 
 \bibitem{Muk07} A. Mukherjee, D.J. Hinde, M. Dasgupta et al., Phys. Rev. C \textbf{75}, 044608 (2007). 
 \bibitem{Vio62}  Victor E. Viola, Torbjorn Sikkeland, Phys. Rev. \textbf{128}, 767 (1962). 
 \bibitem{Lei95}  J.R. Leigh, M. Dasgupta, D.J. Hinde et al., 
 Phys. Rev. C \textbf{52},  3151 (1995). 
 \bibitem{Zhang94} H. Q. Zhang, Z. H. Liu, J. C. Xu et al.,  Phys. Rev. C \textbf{49}, 926 (1994). 
 \bibitem{Gom94} P.R.S. Gomes, I.C. Charret, R. Wanis et al., 
 Phys. Re. C \textbf{49},  245 (1994). 
 \bibitem{Hind07} D. J. Hinde, M. Dasgupta, N. Herrald et al., Phys. Rev. C \textbf{75}, 054603 (2007). 
 \bibitem{Frei87} R. Freifelder, P. Braun–Munzinger, P. DeYoung et al., Phys. Rev. C \textbf{35}, 2097 (1987). 
 \bibitem{Kola12} J.J. Kolata, A. Roberts, A.M. Howard et al., Phys. Rev. C \textbf{85}, 054603 (2012). 
 \bibitem{Trot05} M. Trotta, A.M. Stefanini, S. Beghini et al., 
 Euro. Phys. J. A \textbf{25},  615 (2005). 
 \bibitem{Prok08} E.V. Prokhorova, A.A. Bogachev, M.G. Itkis et al., Nucl. Phys. A \textbf{802}, 45 (2008). 
 \bibitem{Stef90} A.M. Stefanini, L. Corradi, H. Moreno et al., Phys. Lett. B \textbf{252}, 43 (1990). 
 \bibitem{Wolf89} F.L.H. Wolfs, R.V.F. Janssens, R. Holzmann et al., Phys. Rev. C \textbf{39}, 865 (1989). 
 \bibitem{Liang07} J. F. Liang, D. Shapira, J. R. Beene et al., Phys. Rev. C \textbf{75}, 054607 (2007).
 
 \bibitem{Koz16} E. M. Kozulin, G. N. Knyazheva, K. V. Novikov, et al., Phys. Rev. C \textbf{94}, 054613  (2016).

 \bibitem{Mis07} S. Misicu and H. Esbensen, Phys. Rev. Lett. \textbf{96}, 112701 (2006) .
 
 \bibitem{Nhu21} N. Nhu Le, N. Ngoc Duy, N. Quang Hung, Eur. Phys. J. A \textbf{57}, 187 (2021). 


 \bibitem{wangbing21} X.-J. Lv, Z. Y. Yue, W.-J. Zhao, B. Wang, Phys. Rev. C \textbf{103}, 064616 (2021).

 \bibitem{Ogan24} Y. Oganessian, Eur. Phys. J. A \textbf{60}, 227 (2024).
 
 \bibitem{EBD} H. L\"u, A. Marchix, Y. Abe, D. Boilley, Comp. Phys. Comm. \textbf{200}, 381 (2016).

 
 \bibitem{Mei} B. Mei, D. L. Balabanski, et al., Chin. Phys. C \textbf{45}, 054001 (2021) 

    \end{thebibliography}
\end{document}